# Linear-Scaling Quantum Transport from Machine-Learning Density Functional Theory Hamiltonians

Bang Liu[1], Yang Zhong[1], Zhi-Xin Guo[2‡], Xin-Gao Gong[1] and Hongjun Xiang[1†]

[1]*Key Laboratory of Computational Physical Sciences (Ministry of Education), Institute of Computational Physical Sciences, State Key Laboratory of Surface Physics, and Department of Physics, Fudan University, Shanghai 200433, China.*

[2]*State Key Laboratory for Mechanical Behavior of Materials, Xi'an Jiaotong University, Xi'an, Shaanxi, 710049, China.*

[‡]zxguo08@xjtu.edu.cn

[†]hxiang@fudan.edu.cn

## Abstract:

Quantum transport simulations that combine density functional theory (DFT) with the nonequilibrium Green's function formalism (DFT-NEGF) are important to modern technology, yet their unfavorable scaling has long confined predictive simulations to small, idealized systems far below the ten-thousand-atom scale of realistic devices. Here, we introduce HamGNN-NEGF, a linear-scaling framework with DFT-level fidelity. An E(3)-equivariant graph neural network trained on conventional DFT Hamiltonians of small structures predicts Hamiltonians for large devices, avoiding costly DFT-NEGF training data. The predicted Hamiltonians are integrated with DFT-derived electrode self-energies, a nonorthogonal kernel polynomial method for Fermi-level determination, and a recursive Green's function algorithm, yielding a computational cost that scales linearly with device length at fixed cross section. Even for devices containing fewer than 500 atoms, HamGNN-NEGF achieves speedups exceeding three orders of magnitude over fully self-consistent DFT-NEGF, with the advantage increasing further with system size. Benchmarks on pristine Pt–Si–Pt, doped Pt–Si:P–Pt, and Pt–molecule–Pt junctions demonstrate meV-level Hamiltonian accuracy, faithful transmission spectra, and predictive simulations beyond 10,000 atoms. Eliminating transport self-consistency also enables hybrid functionals such as HSE06 without additional NEGF overhead, while a zero-bias Hamiltonian approximation extends the framework to finite-bias transport in weakly nonlinear regimes. HamGNN-NEGF thus bridges first-principles accuracy and device-scale simulation, providing a practical route toward predictive modeling of realistic nanoelectronic and quantum devices.

## Introduction

Commercial transistor technologies have entered the sub-10-nm regime, where atomistic device models may contain tens of thousands of atoms[1,2]. At this scale, device performance is strongly influenced by quantum confinement and tunneling, the discrete distribution of dopants and defects, and atomistic phenomena such as structural reconstruction at surfaces and interfaces[3-5]. Conventional technology computer-aided design (TCAD) approaches[6,7], which rely largely on empirical parameters and continuum transport models, struggle to capture these inherently atomistic and quantum-mechanical effects. Predicting the performance of advanced semiconductor devices therefore increasingly requires parameter-free quantum transport simulations that operate directly at the atomic scale.

Density functional theory combined with the nonequilibrium Green's function formalism (DFT-NEGF) provides a rigorous first-principles framework for meeting this need, treating the device electronic structure and its coupling to semi-infinite electrodes on an equal footing and directly yielding transmission spectra and conductance[8-14]. Its application to realistic devices, however, remains severely constrained by computational cost. In conventional fully self-consistent DFT-NEGF calculations, two major operations introduce $O(N^3)$ scaling with the matrix dimension N: explicit diagonalization for Fermi-level determination and repeated solution or inversion of energy-dependent Green's-function matrices during the open-boundary self-consistent-field iterations[14-18]. Fully self-consistent DFT–NEGF (hereafter referred to as Full-NEGF) has therefore remained largely confined to relatively small, idealized systems, far below the ten-thousand-atom scale relevant to realistic devices[19,20]. Consequently, key mesoscopic features, including extended doped channels[21,22], heterogeneous junctions[23], and structurally complex or defective contacts[24-26], remain difficult to address using conventional first-principles transport methods. Bridging this scale gap is thus a central challenge in computational nanoelectronics.

Machine learning (ML) may pave a new way to overcome this bottleneck. Equivariant neural-network models now predict *ab initio* Hamiltonians directly from atomic structure at near-DFT accuracy and have reshaped bulk band-structure calculations[27-30]. Existing data-driven strategies, including direct prediction of observables such as transmission spectra and current-voltage characteristics[31-33], learning of intermediate electrostatic potentials[34,35], and empirical tight-binding parametrizations[12,13], either forgo the rigorous *ab initio* Hamiltonian structure or suffer from limited transferability under changing boundary conditions. In addition, the high cost of achieving open-boundary self-consistency has so far confined most ML-accelerated NEGF implementations to small- and medium-sized systems and semilocal exchange-correlation functionals. These limitations underscore the urgent need for a transferable framework that preserves the full first-principles NEGF formalism, scales efficiently to mesoscopic regimes (>10,000 atoms), and remains structurally compatible with high-accuracy hybrid functionals such as HSE06[36].

In this work, we introduce HamGNN-NEGF, a framework enabling linear-scaling quantum transport with *ab initio* accuracy. An E(3)-equivariant graph neural network (HamGNN), trained on small structures, predicts the large-scale device Hamiltonian directly from its atomic configuration. The predicted Hamiltonians are incorporated into a DFT-Hamiltonian-based NEGF formalism (hereafter referred to as DFTH-NEGF) which evaluates transmission using a static DFT Hamiltonian and electrode self-energies evaluated from bulk DFT calculations, thereby eliminating computationally prohibitive self-consistent transport iterations. Combined with a non-orthogonal kernel polynomial method (KPM) for determining the Fermi level and the recursive Green's function (RGF) algorithm, the complete workflow scales linearly with device length at a fixed cross section. For devices containing 320-480 atoms, HamGNN-NEGF achieves speedups of approximately 1,000-2,000-fold over conventional Full-NEGF, with the advantage increasing further with system size. The framework attains meV-level accuracy across Pt–Si–Pt, doped Pt–Si:P–Pt, and Pt–molecule–Pt junctions, extending DFT-level quantum transport simulations to the 10,000-atom regime. Moreover, eliminating self-consistency permits adopting advanced exchange-correlation treatments (such as the HSE06 hybrid functional, essential for semiconductor alignments) at zero additional transport cost, unlocking high-fidelity mesoscopic simulations previously deemed computationally inaccessible.

## Results

To clarify the computational origin of this advantage, we first contrast HamGNN-NEGF with the conventional Full-NEGF workflow. As illustrated in Fig. 1(a), conventional Full-NEGF calculations couple the electronic-structure and transport solvers through an iterative open-boundary self-consistent field (SCF) loop, leading to substantial computational cost. This workflow suffers from several major computational bottlenecks. First, determining the device Fermi level necessitates an explicit matrix diagonalization, introducing an inherent $O(N^3)$ scaling burden with the matrix dimension ($N$). Second, in a direct dense-matrix implementation, evaluating the retarded Green's function for the device region requires solving or inverting the energy-dependent Green's-function matrix, whose cost scales as $O(N^3)$. Third, this operation must be performed at numerous energy points along the equilibrium contour and, under finite bias, within the nonequilibrium energy window. It must also be repeated at every SCF iteration to update the density matrix $\boldsymbol{\rho}$ and charge density $\boldsymbol{n}(\boldsymbol{r})$. This cycle continues until full self-consistency is reached before the final transmission spectrum ($T$(E)) is evaluated, severely limiting the accessible system sizes in traditional Full-NEGF calculations.

HamGNN-NEGF addresses these limitations by targeting the two key challenges in large-scale first-principles transport: efficient construction of DFT-level device Hamiltonians and Fermi-level determination in large non-orthogonal systems. As shown in Fig. 1(b), the framework removes these bottlenecks through two complementary strategies. First, it bypasses the SCF cycle through a decoupled DFTH-NEGF approach, replacing the open-boundary self-consistent Hamiltonian with a conventional periodic DFT Hamiltonian coupled to electrode self-energies. Second, it eliminates the diagonalization bottleneck by generalizing the KPM[37-39] to the non-orthogonal case for Fermi-level determination.

The HamGNN network is trained on small DFT-computed periodic device structures to learn the mapping from atomic configurations to these periodic Hamiltonian matrices. Once trained, it predicts the Hamiltonian $H_{ML}$ of an arbitrarily large device in a single forward pass, supplying the transport backend while entirely avoiding iterative self-consistency. Given the predicted Hamiltonian in the non-orthogonal basis, we employ the non-orthogonal KPM to obtain the Fermi level without explicit diagonalization. In a non-orthogonal localized basis, the Kohn-Sham equation is formulated as a generalized eigenvalue problem, $H_c\boldsymbol{c}_i = \varepsilon_i S_c \boldsymbol{c}_i$, where $H_c$ and $S_c$ are the Hamiltonian and overlap matrices of the central scattering region, respectively. $\varepsilon_i$ is the eigenvalue of the $i$-th state, and $\boldsymbol{c}_i$ is the corresponding eigenvector. We therefore expand the rescaled operator $\widetilde{A} = S_c^{-1}(H - bS_c)/a$ in Chebyshev polynomials. The action of $\widetilde{A}$ on a vector is evaluated implicitly by solving linear systems involving the overlap matrix $S_c$, rather than by explicitly constructing $S_c^{-1}$. The resulting Chebyshev moments reconstruct the density of states to determine the Fermi level via the electron-number constraint[38,39]. Finally, the transmission spectrum is evaluated using the efficient RGF method. By replacing the costly iterative construction of $H_{KS}$ and avoiding explicit diagonalization, this combined workflow achieves substantial computational efficiency.

This efficiency gain is further reflected in the favorable scaling of each component of HamGNN-NEGF. The HamGNN prediction step scales linearly, $O(N)$, under a fixed local cutoff because the number of graph edges grows proportionally with the number of atoms. The non-orthogonal KPM determines the Fermi level with a cost of $O(N_{\mathrm{mom}}N_{\mathrm{rand}}N)$ for sparse matrices, where $N_{\mathrm{mom}}$ is the number of Chebyshev moments and $N_{\mathrm{rand}}$ is the number of stochastic random vectors. The subsequent RGF calculation avoids full matrix inversion and scales as $\mathrm{O}(N_{\mathrm{z}}M_b^3)$ per energy point and per transverse *k*-point, where $N_{\mathrm{z}}$ is the number of principal layers and $M_b$ denotes the matrix dimension of each principal-layer block. Thus, at a fixed transverse cross section, the complete framework scales linearly with device length.

The above workflow rests on a central physical premise: the Hamiltonian can be constructed without explicitly accounting for the open-boundary environment, such that the non-self-consistent DFTH-NEGF strategy provides an accurate approximation to Full-NEGF. We therefore first validate the omission of the SCF loop before deploying the full HamGNN-NEGF framework. Figure 2 compares the transmission spectra obtained from this DFTH-NEGF strategy against Full-NEGF calculations. The benchmarks cover two representative scattering scenarios: an Au-benzenedithiol-Au molecular junction featuring heterogeneous interfaces and charge transfer (Fig. 2(a)), and a continuous one-dimensional carbon nanotube driven by localized Stone-Wales defect scattering (Fig. 2(c)). In both cases, the DFTH-NEGF approach reproduces the rigorous Full-NEGF result almost exactly (Figs. 2(b) and 2(d)). Eliminating the SCF loop also delivers immediate computational savings: relative to the Full-NEGF baseline, DFTH-NEGF reduces the total computational time by 18.3% for the Au–benzenedithiol–Au junction and 27.4% for the carbon nanotube. Given the well-known difficulty of achieving SCF convergence in large-scale heterogeneous junctions, the computational advantage would be much more pronounced as system size and complexity increase, as demonstrated below for the Pt–Si–Pt devices. With the transport methodology thus validated, we proceed to apply the integrated HamGNN-NEGF framework to large-scale device simulations.

We further benchmark the framework on three representative systems with distinct transport mechanisms and chemical environments: a pristine Pt–Si–Pt metal-semiconductor device (Supplementary Fig. 1), a phosphorus-doped Pt–Si:P–Pt device (Supplementary Fig. 2), and a Pt–molecule–Pt junction (Supplementary Fig. 3). Together, they probe whether the learned Hamiltonian can reproduce inorganic metal-semiconductor transport, impurity scattering, and molecular resonant tunneling within the NEGF formalism.

We begin with the pristine Pt–Si–Pt device. As shown in Fig. 3(a), this system consists of semi-infinite Pt electrodes connected to a supercell Si scattering region, where transport is strongly influenced by the metal-semiconductor interface and the electronic structure of the Si channel. Since the Fermi level determined by KPM is used directly in the subsequent NEGF transport calculation, we first examine its numerical convergence with respect to the Chebyshev expansion order $N_{\mathrm{mom}}$ and the number of random vectors $N_{\mathrm{rand}}$. As shown in Supplementary Table 1, the non-orthogonal KPM achieves meV-level accuracy for the Fermi level after convergence, confirming its reliability for the 2×2×5 Si supercell.

Having established the accuracy of the Fermi-level determination, we next assess the complete HamGNN-NEGF workflow. Trained on DFT Hamiltonians of small devices (1×1×2, 1×1×3, and 1×1×4 Si supercells in the scattering region), HamGNN predicts DFT-level Hamiltonian blocks for larger scattering regions directly from atomic structure, with a test-set matrix-element mean absolute error (MAE) of 0.020 meV (Supplementary Fig. 4(a)). This element-wise accuracy provides the basis for using the predicted Hamiltonian in subsequent transport calculations. Specifically, the HamGNN-predicted Hamiltonian for the device containing the 2×2×5 Si supercell is then combined with the electrode self-energies, and the zero-bias transmission spectrum is calculated via the NEGF formalism. As shown in Fig. 3(b), the HamGNN-NEGF transmission spectrum closely follows the DFTH-NEGF reference, particularly around the Fermi level, accurately reproducing the main transmission features with a transmission MAE of 1.71%. This result demonstrates that the learned Hamiltonian meticulously preserves the electronic-structure information required for metal-semiconductor quantum transport.

To assess the scalability of the proposed workflow, we also apply HamGNN-NEGF to Pt–Si–Pt devices with increasingly long Si scattering regions while keeping the transverse cross section fixed. The transmission spectra of these extended systems (Supplementary Fig. 4(b), (c)) follow the DFTH-NEGF references closely, with transmission MAEs of 3.86% and 6.25% for the 2×2×7 and 2×2×10 devices, respectively, well beyond the 1×1×2–1×1×4 supercells used for training. Figure 3(c) further shows the CPU time of the complete HamGNN-NEGF workflow as a function of system size. The total cost can be decomposed into three components: HamGNN Hamiltonian prediction, non-orthogonal KPM Fermi-level calculation, and RGF transmission evaluation. It is seen that the HamGNN prediction step accounts for only a small fraction of the total runtime, whereas the dominant computational cost arises from the KPM and RGF solvers. Note that the computational costs of the non-orthogonal KPM, the RGF step, and the overall workflow all increase linearly with the number of atoms. This favorable scalability enables HamGNN-NEGF to push first-principles quantum transport to devices exceeding 10,000 atoms without sacrificing traditional DFT Hamiltonian accuracy.

Having established its scalability in pristine metal-semiconductor devices, we next examine the transferability of HamGNN-NEGF in chemically perturbed semiconductor environments. To this end, we consider phosphorus-doped Pt–Si:P–Pt devices, where substitutional P dopants introduce localized chemical perturbations into the Si channel (Supplementary Fig. 5(a)). This system provides a stringent test of the machine-learned Hamiltonian in doped semiconductor environments. For a short device accessible to DFT, HamGNN-NEGF closely reproduces the DFT reference, both at the level of Hamiltonian matrix elements, with an MAE of 0.02 meV (Supplementary Fig. 5(b)), and in the zero-bias transmission spectrum, with an MAE of 0.72% in 1×1×5 channel (Supplementary Fig. 5(c)). The agreement persists for longer 1×1×7 and 1×1×10 channels, with transmission MAEs of 2.03% and 1.92% (Supplementary Fig. 6).

We then apply the validated workflow to longer Pt–Si:P–Pt devices beyond the reach of Full-NEGF and evaluate the length dependence of the linear response current at a bias of 0.1 V (Fig. 3(d)). The current decreases overall as the P-doped region is lengthened, while shorter devices exhibit nonmonotonic fluctuations arising from quantum interference. This trend indicates the progressive suppression of transport by impurity scattering. These length-dependent results demonstrate that HamGNN-NEGF provides a practical and transferable tool for investigating transport in large-scale doped semiconductor devices that are otherwise intractable with conventional first-principles methods.

To further verify the accuracy of our model, we evaluated the zero-bias transmission spectra of Pt-Si-Pt and Pt-Si:P-Pt devices across varying lengths. HamGNN-NEGF faithfully captures the Full-NEGF results with minimal MAEs of 2.80-3.61% and 1.00-1.98%, respectively (Supplementary Fig. 7). Beyond yielding high-fidelity spectra, HamGNN-NEGF exhibits substantially improved computational scalability. With the increase in the number of atoms, the computational advantage over Full-NEGF becomes even more significant, achieving a computational speedup of three orders of magnitude in Pt-Si-Pt device (approximately 1,000 to 2,000 times), as shown in Supplementary Fig. 8. The markedly different scaling behaviors of the two approaches suggest that the computational acceleration afforded by HamGNN-NEGF will become even greater for systems containing larger numbers of atoms.

Beyond extended inorganic channels, HamGNN-NEGF can also handle chemically complex contact interfaces, as exemplified by Pt–molecule–Pt junctions (Supplementary Fig. 9(a)). Quantum transport in such molecular devices represents a fundamentally distinct regime, governed by intricate molecule-electrode hybridization and sensitive molecular orbital alignments rather than bulk defect scattering. The framework captures these subtle interfacial chemical environments: the HamGNN-predicted Hamiltonian reproduces the DFT reference to 0.02 meV (Supplementary Fig. 9(b)) and yields a transmission spectrum in close agreement with the DFTH-NEGF reference for a five-ring junction (MAE 2.50%, Supplementary Fig. 9(c)). For longer chains the pointwise error grows to 8.01% and 11.10% for ten- and sixteen-ring junctions (Supplementary Fig. 10). This increase reflects the sensitivity of narrow, densely spaced resonances to small shifts in molecular-orbital energies and molecule-electrode coupling. Nevertheless, the principal resonance positions and the low-transmission region remain well preserved (Supplementary Fig. 10). These results confirm that the data-driven methodology is not limited to periodic or weakly perturbed inorganic lattices but transfers robustly to chemically diverse and structurally complex heterointerfaces.

Another important advantage of HamGNN-NEGF is that it is not intrinsically tied to PBE-level Hamiltonians[40,41]. In principle, the model can be trained on data from any exchange-correlation functional, given consistent reference data. This is particularly relevant for semiconductor devices, where hybrid functionals such as HSE06[36] give improved band gaps and band alignments but are rarely used in Full-NEGF because of their high cost. To demonstrate this capability, we train HamGNN on an HSE06 Hamiltonian dataset for a defective graphene transport device (Supplementary Fig. 11). As shown in Fig. 4(a), the device consists of semi-infinite graphene electrodes connected to a graphene scattering region containing vacancy defects. The vacancy introduces a local perturbation to the $\pi$-conjugated carbon network and acts as a scattering center, providing a stringent test of whether the learned Hamiltonian can reproduce the defect-modified electronic structure at the hybrid-functional level.

As shown in Fig. 4(b), the HamGNN-predicted Hamiltonian matrix elements are in close agreement with the HSE06 reference, yielding an MAE of 0.075 meV on the test set. The corresponding HamGNN-NEGF transmission spectrum with four vacancy defects in the device closely follows the HSE06-level reference result (Supplementary Fig. 12), with a transmission MAE of 0.76%. The approach retains this accuracy at higher defect densities, with transmission MAEs of 1.09% and 1.00% for six- and eight-vacancy devices at the HSE06 level (Supplementary Fig. 13). Applying the identical protocol at the PBE level yields a Hamiltonian MAE of 0.04 meV and transmission MAEs of 0.41%, 0.34% and 0.37% for four-, six- and eight-vacancy devices (Supplementary Figs. 14 and 15), so that both functional levels used in the comparison below are validated to the same standard.

Building upon this validated accuracy and scalability, we next investigate the large-scale transport physics of the defective graphene device at both the PBE and HSE06 levels. Figure 4(c) shows the length-dependent linear response current at a bias of 0.1 V for the two functionals. In the short-channel limit, the current exhibits pronounced nonmonotonic oscillations, a hallmark of phase-coherent resonant tunneling driven by the hybridization of localized defect states with the pristine graphene continuum[42,43]. As the scattering region lengthens, cumulative elastic backscattering effectively suppresses this interference, producing a continuous macroscopic current attenuation that signals a transition toward localized transport[44].

The PBE and HSE06 calculations yield clearly distinguishable current profiles, with the largest differences appearing in the short-channel regime, where resonance peaks and dips shift in both position and amplitude. For the longest devices the two currents gradually converge as incoherent attenuation takes over. These short-channel differences originate from the exact-exchange component of HSE06, which reduces self-interaction errors, suppresses the spurious overdelocalization of defect-induced scattering states, and thereby modifies their dispersion and energetic alignment relative to the transport window[36]. Because the defective graphene channel is metallic, the PBE-HSE06 discrepancy arises mainly from differences in band dispersion rather than from band-gap opening. Naively, one would expect that the HSE current is smaller than the PBE result as HSE predicts more localized states and a larger band gap. Unexpectedly, the HSE current is even larger than the PBE case for several different device lengths. This unexpected behavior highlights the importance of hybrid-functional accuracy in quantum transport simulations and underscores the value of HamGNN-NEGF, which enables such accuracy at essentially no additional transport cost.

## Discussion

Beyond the zero-bias regime, we extended the HamGNN-NEGF framework to finite-bias transport by employing a zero-bias Hamiltonian (H0) approximation, in which the Hamiltonian obtained at zero bias is supplied directly to the NEGF solver and only the electrode electrochemical potentials and the nonequilibrium energy window are updated with bias (Supplementary Section VIII). This approach is particularly well suited to weakly nonlinear systems, in which the low-bias current-voltage characteristics deviate only slightly from Ohm's law and bias-induced changes in the electrostatic potential and electronic structure remain small.

We first isolated the error introduced by the H0 approximation itself, independently of the machine-learned Hamiltonian, using a carbon nanotube containing a Stone-Wales defect (Fig. 2(c)). Over the range 0.2-1.0 V, the transmission spectra obtained from H0-based NEGF (H0-NEGF) and from fully self-consistent finite-bias Full-NEGF remain closely superimposed, deviating appreciably only above ~1 eV at the two highest biases (Supplementary Fig. 16(a)-(e)). Such deviations lie outside the bias window over which the current is integrated, so that the current-voltage characteristics are essentially indistinguishable (Supplementary Fig. 16(f)): the absolute current differences are 0.067, 1.093, 0.479, 1.203 and 0.014 μA at 0.2, 0.4, 0.6, 0.8 and 1.0 V, corresponding to relative deviations below 2% across the entire range. The H0 approximation therefore preserves the transport observable of practical interest even where the full-window transmission error grows.

Applying the complete HamGNN-NEGF workflow at finite bias, so that the errors of the H0 approximation and of the ML Hamiltonian act together, we obtain transmission spectra in good agreement with Full-NEGF for both undoped Pt–Si–Pt and phosphorus-doped Pt–Si:P–Pt junctions across the supercell sizes examined (Supplementary Figs. 17-19 and 23-25). The agreement is markedly better for the doped junctions, where the transmission MAE remains within 1.5% at both 0.1 and 0.2 V, than for the pristine metal-semiconductor junctions, where it grows from 4-5% at 0.1 V to 8-13% at 0.2 V because their dense, narrow interface-derived transmission features are far more sensitive to the small bias-induced shifts that the H0 approximation does not capture. The spatially resolved local density of states (LDOS) at 0.1 V is likewise reproduced throughout the device, including the depleted region that spans the semiconductor channel and its narrowing at the two metal contacts (Supplementary Figs. 20-22 and 26-28). Although the H0 approximation is expected to break down in strongly nonlinear regimes such as those exhibiting negative differential resistance (NDR), these results establish HamGNN-NEGF as a reliable and computationally efficient alternative for simulating finite-bias transport in weakly nonlinear devices.

In summary, we have introduced HamGNN-NEGF, a scalable first-principles quantum transport framework that bypasses the cubic-scaling bottlenecks of Full-NEGF calculations. Using an E(3)-equivariant graph neural network trained on small periodic structures, the framework predicts DFT-level Hamiltonians for large devices and directly supplies them to a DFTH-NEGF transport solver. Combined with non-orthogonal KPM for Fermi-level determination and RGF for transmission evaluation, the workflow achieves linear scaling of quantum transport computational cost with device length. For the benchmark devices containing 320-480 atoms, HamGNN-NEGF delivers speedups of approximately 1,000-2,000 times, with the computational advantage expanding massively as the atomic count increases. Extensive benchmarks on metal-semiconductor interfaces, doped channels, molecular junctions, and defective graphene demonstrate meV-level Hamiltonian accuracy, faithful reproduction of scattering features, and predictive simulations for devices exceeding 10,000 atoms. Furthermore, advanced functionals such as HSE06 can be incorporated with virtually no extra transport overhead, narrowing the gap between ab initio accuracy and mesoscopic device dimensions. A zero-bias Hamiltonian approximation further extends the framework to finite-bias transport in weakly nonlinear regimes at minimal additional cost. Looking ahead, extending HamGNN-NEGF to strongly nonlinear regimes (e.g., negative differential resistance) via a Δ-learning framework for bias-dependent Hamiltonians represents a natural next step, for which this work provides a solid foundation.

# Method

The main DFT datasets were generated at the generalized-gradient approximation (GGA) level using the Perdew-Burke-Ernzerhof (PBE) exchange-correlation functional[40], as implemented in the SIESTA package[18]. Norm-conserving pseudopotentials and a double-ζ plus polarization (DZP) basis set were used. Three classes of devices were considered: Pt–Si–Pt metal-semiconductor devices, Pt–molecule–Pt molecular junctions, and P-doped Pt–Si:P–Pt devices. For each class, three scattering-region lengths were constructed, and 50 configurations were generated for each length by applying random atomic perturbations of 0.08 Å to the reference structure in the scattering region. The Hamiltonian and overlap matrices were extracted from ground-state DFT calculations and used as reference data for HamGNN training. The datasets were divided into training, validation, and test sets with a ratio of 0.8:0.1:0.1. The k-point meshes[45] were chosen according to the periodicity and transverse size of each device and were set to [11×11×1], [9×9×1], and [11×11×1] for Pt–Si–Pt, Pt–molecule–Pt, and Pt–Si:P–Pt in datasets, respectively. For the benchmark systems, the Fermi level was obtained by direct diagonalization of the generalized eigenvalue problem. Zero-bias transmission spectra were calculated using a DFTH-NEGF procedure, in which either the DFT Hamiltonian or the HamGNN-predicted Hamiltonian was combined with Pt electrode self-energies in TBtrans package. The Full-NEGF were calculated by TRANSIESTA[14]. For TBtrans and TRANSIESTA calculation, the k-point meshes were set to [7×7×1], [9×9×1], and [11×11×1] for Pt–Si–Pt, Pt–molecule–Pt, and Pt–Si:P–Pt, respectively. The datasets of the Pt–Si–Pt, Pt–molecule–Pt, and Pt–Si:P–Pt devices are shown in Supplementary Figs. 1-3 (Section I of the Supplementary Information).

Additional hybrid HSE06 functional calculations were carried out using the HONPAS package[46,47] to assess the transferability of the Hamiltonian-learning strategy beyond the PBE level. The corresponding HSE06 Hamiltonian and overlap matrices were generated from ground-state calculations and used for separate HamGNN training and validation following the same dataset construction and splitting protocol. The mean absolute error (MAE) calculation of the transmission spectrum is $MAE = 1/N\sum(T_{predict}(E_i) - T_{target}(E_i))$.

The calculated structures of the dataset of defective graphene transport device is shown in Supplementary Fig. 11. The current at a bias voltage $V_b$ can be calculated as

$$I(V_b) = \frac{2e}{h}\int_{-\infty}^{+\infty} dE[f(E,\mu_L) - f(E,\mu_R)]T(E,V_b). \quad (1)$$

Here the $f(E,\mu_L)$ and $f(E,\mu_R)$ are the Fermi-Dirac distribution of the left and right electrodes, respectively. $\mu_L$ and $\mu_R$ are the electrochemical potentials of the left and right electrodes, respectively. The $T(E,V_b)$ is the transmission probability for an electron at energy $E$ at a bias $V_b$. In the linear response method, the $V_b$ in the transmission $T(E,V_b)$ is set to 0 V.

For weakly nonlinear systems under low bias, the finite-bias transport properties were evaluated by directly supplying zero-bias Hamiltonians (H0) predicted by HamGNN into the NEGF solver. This scheme bypasses computationally expensive finite-bias SCF iterations while accurately capturing bias-dependent transmission spectra, currents, and local density of states.

All first-principles calculations and quantum transport simulations were executed on a computational node equipped with Intel(R) Xeon(R) CPU Max 9462 processors, while the training of the HamGNN model was accelerated using a single NVIDIA H100 GPU.

## Acknowledgements

We acknowledge financial support from Natural Science Foundation of China (grants No. 12188101, 12474237, 52371236), the National Key R&D Program of China (No. 2022YFA1402901), Shanghai Science and Technology Program (No. 23JC1400900), the Guangdong Major Project of the Basic and Applied Basic Research (Future functional materials under extreme conditions--2021B0301030005), Shanghai Pilot Program for Basic Research—FuDan University 21TQ1400100 (23TQ017), the robotic AI-Scientist platform of Chinese Academy of Science, and New Cornerstone Science Foundation and Science Fund for Distinguished Young Scholars of Shaanxi Province (No. 2024JC-JCQN-09).

## Author contributions

H. X. and Z.-X. G. proposed the research and the methodology in this work. B. L. wrote the codes, performed the calculation and wrote the paper. Y. Z. checked codes and ML training results. H. X., Z.-X. G. and X.-G.G. revised the paper.

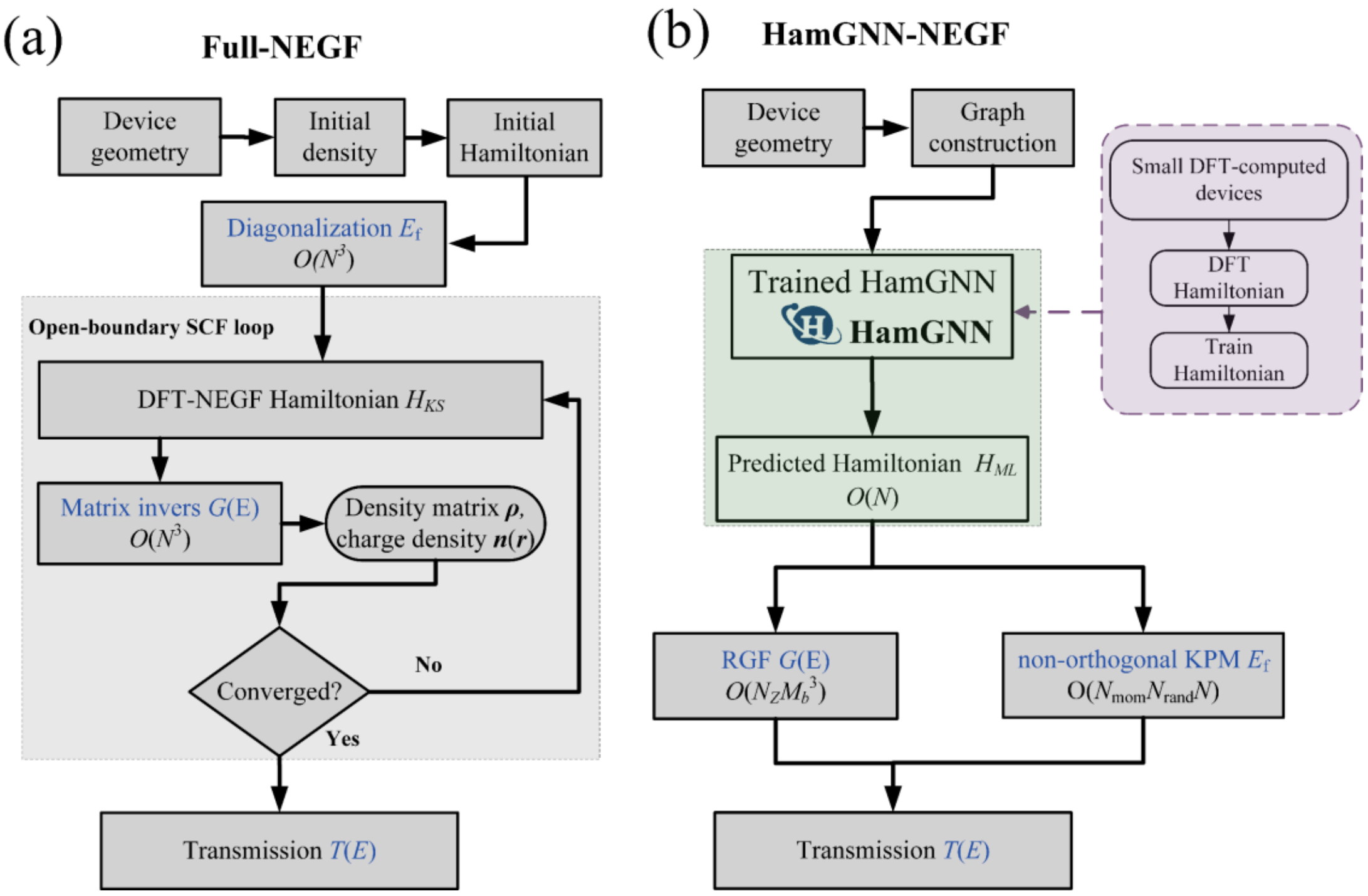

**Figure 1** Workflows of (a) Full-NEGF and (b) HamGNN-NEGF. In (a), after determining the Fermi level $E_f$, the open-boundary SCF loop iteratively evaluates the Green's function $G$(E) to update the density matrix $\boldsymbol{\rho}$, charge density $\boldsymbol{n(r)}$, and Hamiltonian $H_{KS}$ until convergence, before finally computing the transmission $T(E)$. In (b), the trained HamGNN predicts $H_{ML}$ from the device graph, followed by Fermi-level evaluation using non-orthogonal KPM and Green's-function calculation via RGF to obtain $T$(E). Offline HamGNN training uses small DFT-computed devices and Hamiltonians as labels. Here, $N$ is the total system size, $N_z$ is the number of principal layers, and $M_b$ is the corresponding block dimension. For the KPM, $N_{mom}$ and $N_{rand}$ denote the number of Chebyshev moments and random vectors, respectively. Replacing the trained HamGNN in (b) with a conventionally computed DFT Hamiltonian yields the DFTH-NEGF workflow.

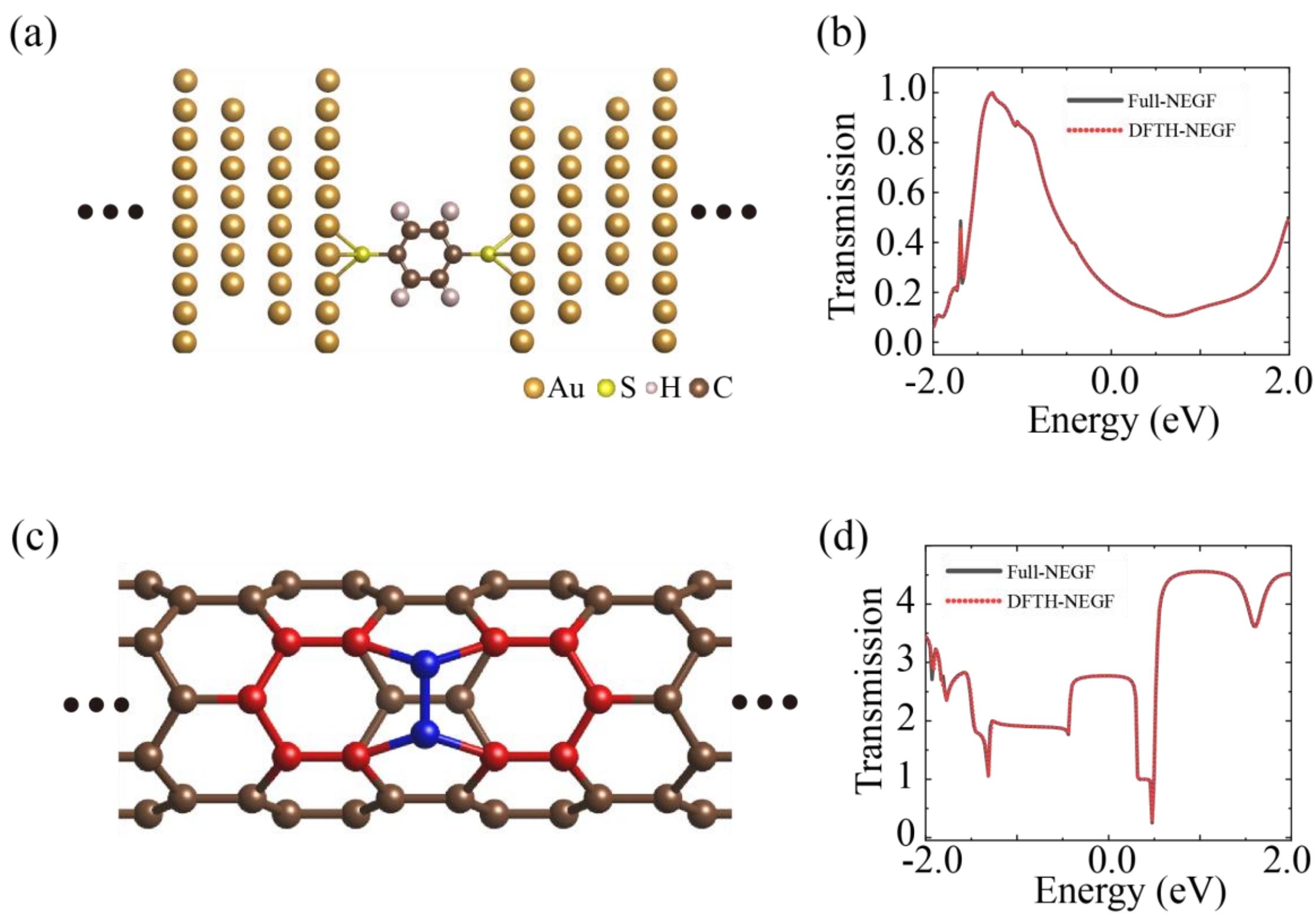

**Figure 2** Comparison of baseline transport calculations from the DFTH-NEGF method with Full-NEGF. Atomic structures of (a) Au–benzenedithiol–Au molecular junction device and (c) a carbon nanotube featuring a Stone-Wales defect, where the red and blue spheres highlight the carbon atoms constituting the topological defect. The corresponding transmission spectra are shown in (b) and (d), respectively.

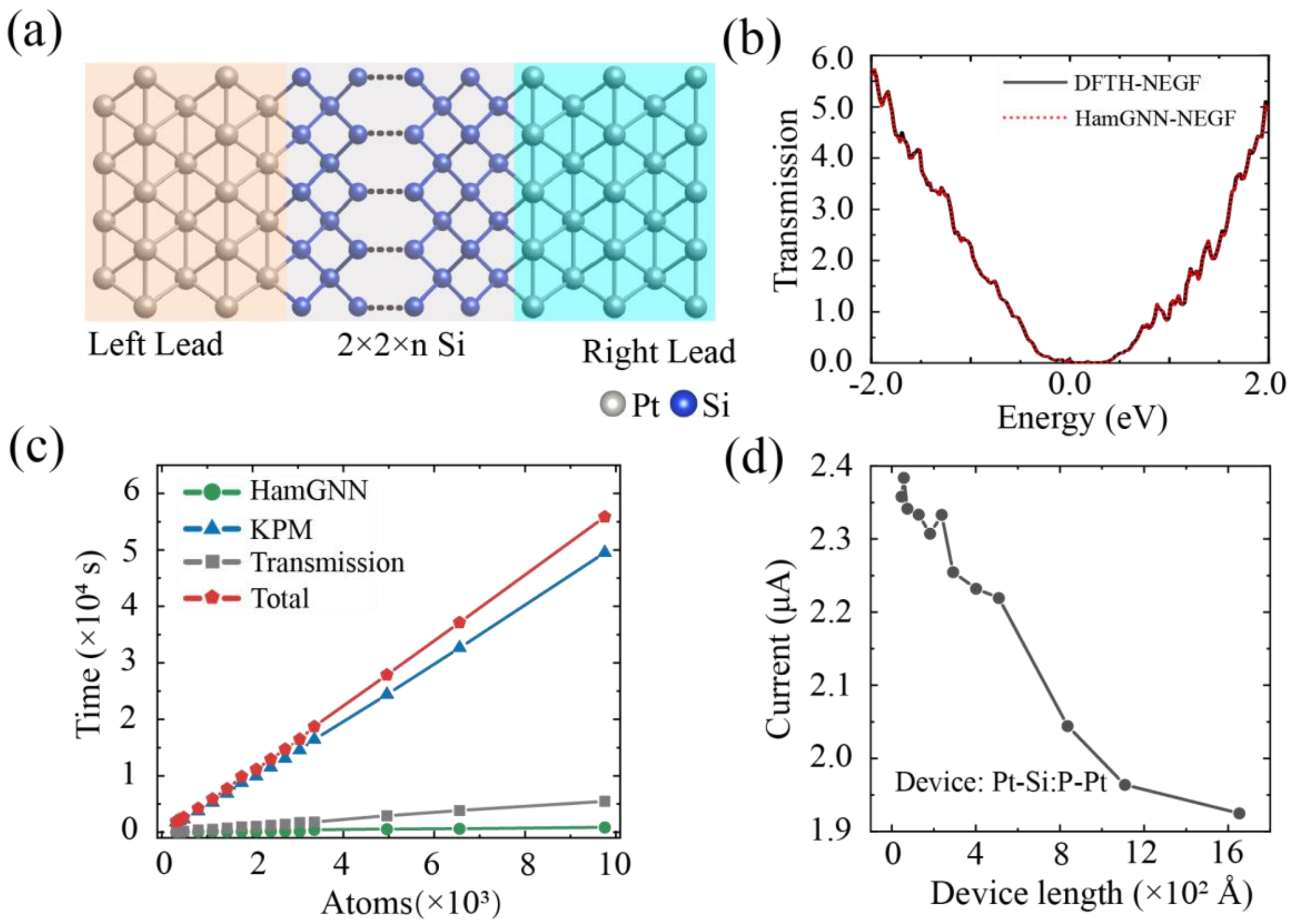

**Figure 3** Validation and computational scaling of the HamGNN-NEGF framework for metal-semiconductor devices. (a) Schematic atomic structure of the Pt–Si–Pt device. (b) The comparison of zero-bias transmission spectra calculated using DFTH-NEGF and HamGNN-NEGF with 2×2×5 Si-supercell in the scattering region of Pt–Si–Pt device. (c) Scaling of the CPU time for the HamGNN-NEGF. The total computational time and its workflow contributions from HamGNN Hamiltonian prediction, KPM Fermi-level calculation, and transmission calculation are shown as a function of system size of Pt–Si–Pt device. Note that file input/output (I/O) overhead is excluded from the reported times. (d) Length dependence of the linear response current in the Pt–Si:P–Pt device at bias of 0.1 V.

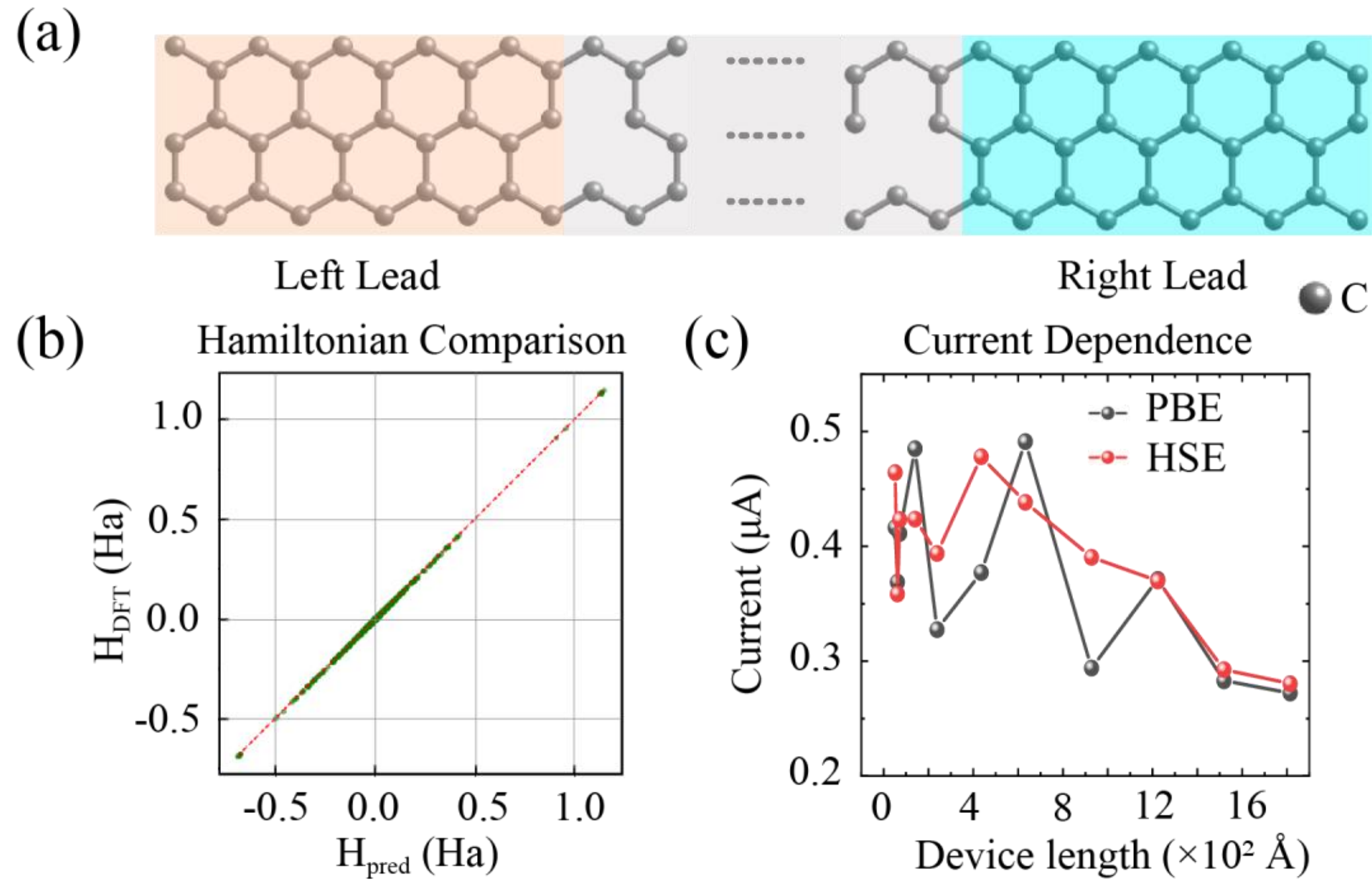

**Figure 4** Application of the HamGNN-NEGF framework to defective graphene at the HSE06 hybrid-functional level. (a) Schematic atomic structure of the defective graphene device. (b) Test-set parity plot comparing the Hamiltonian matrix elements predicted by HamGNN with the corresponding DFT-HSE06 reference values. (c) Device-length dependence of the current at an applied bias of 0.1 V. The corresponding transmission spectra are calculated using HamGNN-NEGF.